\documentclass[english]{article}

\usepackage[T1]{fontenc}
\usepackage[latin9]{inputenc}
\usepackage{geometry}
\usepackage{graphicx}
\usepackage{babel}
\usepackage{hyperref}
\usepackage{color}
\usepackage[textsize=small]{todonotes}
\usepackage{soul} 
\usepackage{natbib}
\usepackage{setspace}
\usepackage{caption}

\bibpunct{[}{]}{,}{n}{}{;}
\newcommand{\edit}[1]{\textcolor{black}{#1}}%

\title{JID - spatial big data opinion}
\begin{document}

\begin{titlepage}
\begin{center}
{\huge Mind the scales: Harnessing spatial big data for infectious disease surveillance and inference}

\vspace{0.3 cm}
{\large Elizabeth C. Lee$^{1,\dag}$ (ecl48@georgetown.edu)\\
Jason M. Asher$^{2}$ (jason.m.asher@gmail.com)\\
Sandra Goldlust$^{1}$ (sandra.goldlust@georgetown.edu)\\
John D. Kraemer$^{3}$ (jdk32@georgetown.edu)\\
Andrew B. Lawson$^{4}$ (lawsonab@musc.edu)\\
Shweta Bansal$^{1,5,*}$ (shweta.bansal@georgetown.edu)}

\vspace{1cm}
\noindent
$^{1}$Department of Biology, Georgetown University, Washington, District of Columbia, United States of America\\
$^{2}$Leidos support to Department of Health and Human Services, Washington, District of Columbia, United States of America\\
$^{3}$Department of Health Systems Administration, Georgetown University, Washington, District of Columbia, United States of America\\
$^{4}$Department of Public Health Sciences, Medical University of South Carolina, Charleston, South Carolina, United States of America\\
$^{5}$Fogarty International Center, National Institutes of Health, Bethesda, Maryland, United States of America\\
$^{*}$Corresponding author; $^{\dag}$Alternate corresponding author

\vspace{1cm}
\end{center}

\section*{Abstract} 
Spatial big data have the ``velocity,'' ``volume,'' and ``variety'' of big data sources and \edit{contain additional geographic information}. Digital data sources, such as medical claims, mobile phone call data records, and geo-tagged tweets, have entered infectious disease epidemiology as novel sources of data to complement traditional infectious disease surveillance. In this work, we provide examples of how spatial big data have been used thus far in epidemiological analyses and describe opportunities for these sources to improve disease mitigation strategies and public health coordination. In addition, we consider the technical, practical, and ethical challenges with the use of spatial big data in infectious disease surveillance and inference. Finally, we discuss the implications of the rising use of spatial big data in epidemiology to health risk communication, and public health policy recommendations and \edit{coordination across scales}. \\

\noindent
\textbf{Keywords: spatial big data, spatial epidemiology, disease mapping, infectious disease, digital epidemiology, statistical bias}\\

\noindent
Abstract word count: 134; Text word count: 2771

\end{titlepage}


\doublespacing


\section*{Introduction}

During one of epidemiology's formative moments, John Snow mapped London households with cholera and succeeded in highlighting the risk associated with the Broad Street pump. Since then, spatial investigations have played a critical role in improving our understanding of the associations between risks and disease outcomes. In infectious disease epidemiology, we ask: \emph{Which populations are at higher risk for disease? Where did this outbreak originate? Where can we expect future disease outbreaks to arise?} \textemdash fundamentally, these are spatial questions that rely on spatial data for answers.

Traditional infectious disease epidemiology is built on the foundation of \edit{relatively} high quality and high accuracy data on disease (e.g., serological diagnostics) and behavior (e.g., \edit{vaccination} surveys). These data are usually characterized by small size, but they benefit from \edit{control groups or designed observational samples from known underlying populations, thus rendering} it possible to make population-level inferences. On the other hand, digital infectious disease epidemiology \edit{typically} uses existing digital traces, re-purposing them to identify patterns in health-related processes. Digital data are electronic in form, and can often be characterized as ``big data'' when they are produced in large volumes (i.e. a large number of subjects or a large number of measurements per subject), with high velocity (i.e. data created in near real-time), and have variety in sources and organizational structures \cite{Laney2001}. When big data are characterized by fine spatial granularity, identifying point or areal locations, we refer to them \edit{here} as \edit{\textit{spatial big data}}. Big data provide opportunities for infectious disease epidemiology and public health because they increase accessibility to populations over space and time; data on personal beliefs, behaviors, and health outcomes are now available at unprecedented breadth and depth. The trade-off to this tremendous access is \edit{the potential for} loss of quality and accuracy. Streams of digital data relevant to public health \edit{may} serve as proxies for a desired measure, \edit{but these datasets may not meet the assumptions for} standard \edit{methods of epidemiological comparison (e.g., self-reported symptoms on social media and serological diagnoses both serve as proxies for ``true cases,'' but they have different biases and collection procedures, and represent different populations)}.   

The trade-off between access and accuracy, and the task of separating true signal from large and varied noise characterizes the challenge and opportunity of big data for infectious disease epidemiology \cite{Khoury2014}. In this piece, we focus specifically on spatial big data and its applications to the field of spatial epidemiology. We highlight the opportunities for spatial big data to improve spatial modeling and data coverage, and describe ongoing challenges as spatial big data become more pervasive in informing disease surveillance, disease control, and public health policy.

\section*{Spatial big data open new doors in epidemiology}

True to the promise of \emph{variety} in big data streams, several familiar technologies produce spatial big data that can be used for infectious disease surveillance and modeling. Social media sites like Facebook and Twitter allow users to tag individual posts with specific locations, linking \edit{geography} to specific health behaviors. Mobile phones send signals with GPS locations and their call data records are spatially referenced through cell tower locations, both of which enable the measurement of human activity and mobility \edit{\cite{Wesolowski2012a, Wesolowski2015a}}. Web search data may capture user location through internet IP addresses, and online encyclopedia (Wikipedia) access logs may identify locations based on the search language \cite{Ginsberg2009, Generous2014}. Administrative medical claims and pharmacy transactions indicate the location of healthcare facilities and drugstores where patients seek care and medications \edit{\cite{Viboud2014, Pivette2014}}. Restaurant reservation cancellations on sites like OpenTable may provide insight into disease incidence in specific cities \cite{Nsoesie2014}.

Infectious disease epidemiology has already witnessed an impact from spatial big data, and the development of new methodologies and improvements to computational efficiency will only increase the potential of these data sources. Satellite imaging to infer climate, land use, and population density information has contributed to a better understanding of the spatial distribution of critical mosquito disease vectors and the seasonal epidemic dynamics of measles \citep[e.g.,][]{Kraemer2015, Bharti2011}. HealthMap, an automated, online news and outbreak reporting aggregator, has enabled the assimilation of disparate sources of disease occurrence data and has been used to examine spatial dynamics of cholera \cite{tuite2011}. Mobile phone call data records have provided insights into human mobility that have informed risk maps, importation potential, and spatial dynamics of dengue and malaria \citep[e.g.,][]{Wesolowski2015a}. Medical claims data have been used to examine spatial heterogeneity in influenza epidemic timing and severity \cite{Gog2014, Lee2015}, while geo-referenced Twitter data has been used to identify spatial anti-vaccination sentiment \cite{Salathe2011}. 

While these studies with spatial big data have leveraged the fine spatial resolution to develop a detailed understanding of disease risk, there remain untapped opportunities with real-time surveillance, large-scale ecological inference, and adaptive disease mitigation strategies. Harnessing disease data from digital sources may enable epidemiological analyses to be performed at finer spatial scales in areas with poor coverage from traditional public health surveillance, and traditional and digital sources of spatial big data may be combined to \edit{account for the bias and gaps in each \citep[e.g.,][]{Shaman2013}}. The assimilation of multiple spatial big data sources through flexible \edit{statistical} modeling methods, and the continuous nature of data streams could enable near real-time dynamic disease mapping and risk mapping in the near future. For example, Bayesian statistical approaches have emerged as tools for merging multi-scale big data sources, incorporating explicit spatial dependencies into maps and models, and providing a framework for joining disease surveillance data across spatial scales while explicitly capturing the variation \edit{in} measurement bias across locations \citep[e.g.,][]{Corberan-Vallet2014}. \edit{Finally, access to multiple spatial scales of data allows one scale with missing observations to ``borrow information'' from a different scale through the addition of contextual effects in modeling inference \cite{Aregay2015b}}.

\section*{Spatial big data presents technical challenges}

While big data offer significant opportunities for epidemiological modeling and analysis, they also present a variety of technical and practical challenges. The measurement of incomplete and unrepresented populations, the lack of consistency and reliability in data over time, and \edit{the need for data and model validation} are broad challenges with big data and statistical analysis that are discussed elsewhere \edit{\citep[e.g.,][]{Lazer2014, Althouse2015}}. Here, we discuss a narrower set of challenges that arise specifically from the spatial nature of big data. 

\subsection*{Spatial coverage and representation} 
\begin{figure}[ht!]
  \caption{\textbf{A)} Spatial big data have spatial biases in the populations they represent. For instance, \edit{as reported by the 2013 American Community Survey,} there is spatial variation in home internet access across the United States, which might affect the populations \edit{generating search query data in Google Trends}. \textbf{B)} With static spatial data (depicted left), individuals (represented with different colors) report case events (points) at fixed locations. For instance, the \edit{two} individuals each visited the same doctor's office with symptoms multiple times (points along the `time' axis), so their events are recorded at the same position along the `space' axis \edit{(see overlapping trajectories in the lower part of the `space' axis)}, while the \edit{teal} individual visited a different doctor's office with symptoms three times in a similar time period. Events from the same individual are connected with a dashed line. With dynamic spatial data (depicted right), events are recorded as individuals move through space. For example, the \edit{dark blue} individual \edit{(see trajectory that begins earliest on the `time' axis)} recorded four events when they tweeted about symptoms at work, at the grocery store, at the pharmacy, and at home, so their case events occur at four different positions along the `space axis. Events occur in time dynamically (as show in this figure), but events may also be aggregated to regular time intervals (e.g., weekly). \textbf{C)} Data at different spatial scales may have different magnitudes and variability in time after adjusting for population size, even if they are derived from the same data source. For instance, we observe time-varying fluctuations and variation in epidemic peak timing and magnitude in the county-level disease data (grey) that are lost in the state-level data (black). \textbf{D)} One possible method to protect privacy is to mask individual-level data by aggregating collected data to larger spatial resolutions. In reality, individuals (black circles) may be connected to other individuals through mobile phone calls (black lines). The publicly released data may be aggregated to the level of neighborhoods (green circles), and the number of calls between individuals from different neighborhoods (green lines) would be represented with different weights (here, depicted with varying thickness according to number of individual calls).}
  \centering
      \includegraphics[width=0.75\textwidth]{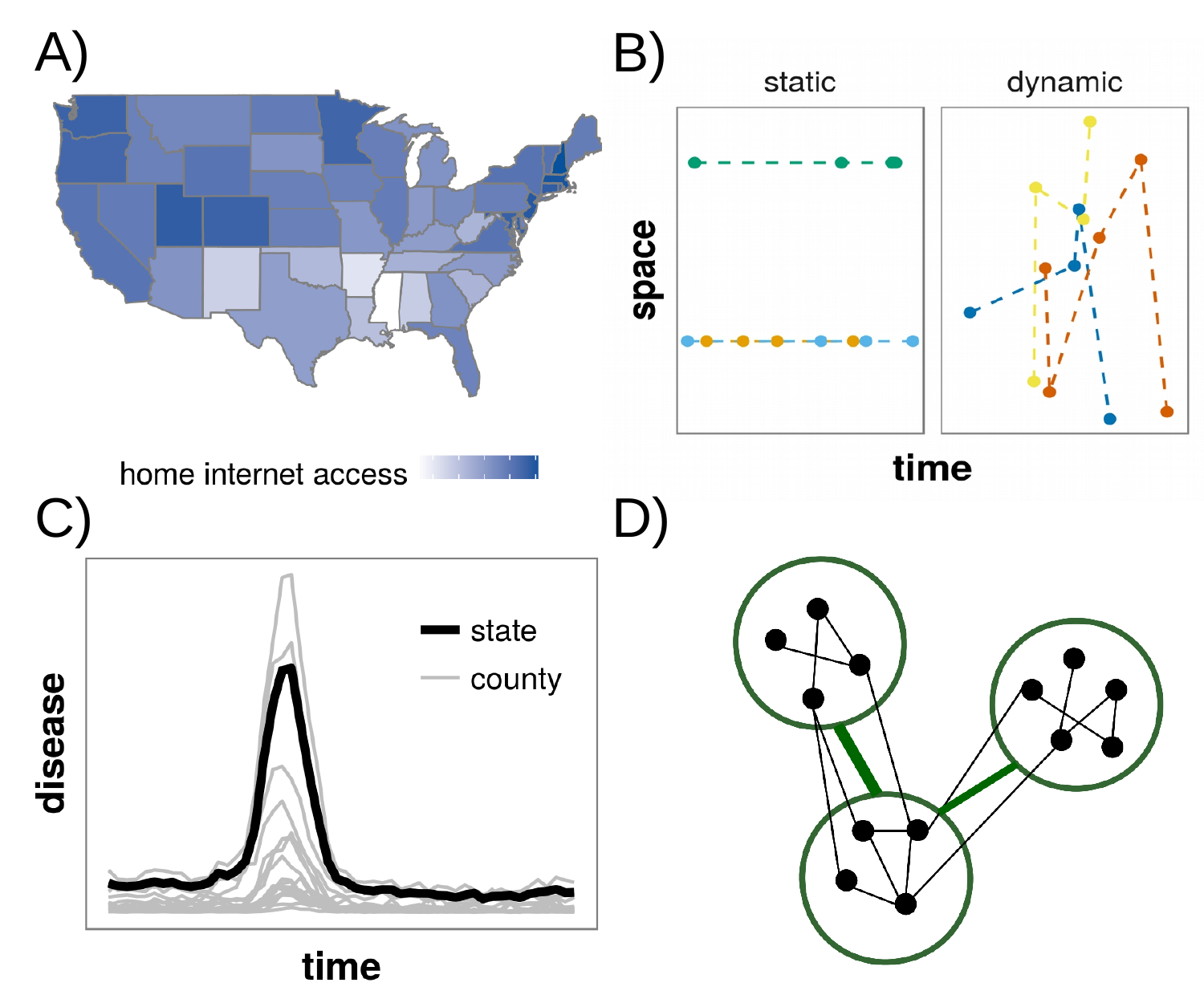}
      \label{fig:fig1}
\end{figure}

Spatial big data may provide precise spatial information, but careful users should question the validity of available data. For one, we know that sources of spatial big data have biases in usership rates and demographics by location (Figure \ref{fig:fig1}A) \edit{\cite{Hecht2014}}. Medical claims record data only from insured and care-seeking populations, which may vary systematically according to socioeconomic and demographic characteristics. Social media sites where users volunteer spatial data tend to have more users and higher quality information per capita in urban areas compared to rural ones \cite{Hecht2014}. Mobile phone \edit{ownership} varies by gender and literacy, and phone sharing between multiple individuals and SIM card switching complicate comparisons of these data across locations \edit{\cite{Wesolowski2012a, Wesolowski2015a}}. \edit{As we cannot often measure the heterogeneities in user populations, these heterogeneities can translate into poor choices in sampling design (e.g., how to stratify samples to get a representative population). Beyond heterogeneities in user populations, the populations captured by big data (e.g., Twitter users) are not usually relevant to epidemiology; even if we could generate an unbiased sample of the population, it may not provide information important to public health. All of these issues complicate} analyses that seek to compare different locations. Ultimately, issues with spatial coverage and representation cause problems for statistical inference, which often depend on assumptions of independent random variation and representative sampling for validity. Future research should compare analyses of spatial big data and analyses of designed observational data in order to demonstrate the \edit{validity} of spatial big data samples and to understand which features of a big data sample can produce \edit{robust} statistical inference.

\subsection*{Spatial uncertainty and noise} 
Each source of big data provides a different type of spatial insight, despite their shared feature of high spatial resolution. Users of social media volunteer their geographical locations in their profiles or \edit{posts}, while internet search engines can log spatial information automatically every time a web search is performed. Sometimes the data are tied to a static location, as in the case of medical claims and health care facilities, but the cell towers associated with call data records and the locations of geo-tagged tweets vary dynamically over time (Figure \ref{fig:fig1}B). Across the combinations of features \textemdash self-reported or automated, static or dynamic \textemdash among these data sources, there are additional layers of uncertainty to consider in the context of epidemiology. \edit{For one, when spatial information in big data is not clearly specified, systematic biases in the results may be generated from the data cleaning process itself (e.g., addresses may be less likely to be geolocated in rural areas) \citep[e.g.,][]{Skelly2002}}. Second, locations of potential transmission events will often differ from locations where disease is reported. While these components are explicitly differentiated in medical claims data (i.e., transmitted in the community, reported at health care facilities), social media posts affiliated with dynamic movements could provide undifferentiated information about both transmission and reporting event locations. Big data provides information at unprecedented levels of spatial precision, but the spatial information fundamental to infectious disease epidemiology (e.g., location and conditions that caused a disease transmission event) continues to remain obscured. As big data becomes more prevalent in epidemiological analysis, public health officials should take care not to conflate spatial precision with spatial accuracy in statistical inference for disease transmission and control.

\subsection*{Spatial scales and misalignment} 
When spatial big data are available at the level of individuals or precise spatial coordinates, practitioners may need to choose the scale of analysis and aggregate data accordingly. Analyzing data at the individual scale is prone to overfitting and the ``atomistic fallacy,'' in which we may make incorrect inferences at the group or population-level based on relationships observed in individual-level data \edit{\cite{Lawson2006}}. For example, if we observe an association between body mass index (BMI) and hospitalization for influenza \edit{among individuals}, it may be incorrect to assume that populations with a high average BMI would have higher rates of influenza-associated hospitalization. On the other hand, analyzing data at aggregated scales is prone to the ``ecological fallacy,'' where inferences about individuals are derived falsely from population-level observations \edit{\cite{Gotway2002, Lawson2006}}. \edit{As an example}, if we observed a negative association between average income and cholera prevalence at a national scale, it would be erroneous to assume that poor individuals have higher risk of cholera than wealthy individuals. Similarly, statistical relationships between predictors and disease outcomes may change when analyses are performed at different spatial aggregations. For instance, Google Flu Trends \edit{attempted} to estimate influenza activity across different regions of the United States by \edit{modeling the relationship between Google search terms and visits for influenza-like illness (ILI)}, as reported in traditional flu surveillance systems \cite{Ginsberg2009}. However, the set of search terms identified as \edit{``most predictive'' of ILI activity were tuned to specific spatial scales}, which means that \edit{terms that fit well at the region-level} may not apply to finer resolution data \edit{\cite{Ginsberg2009, Olson2013}}. Additionally, spatial questions often require the use of multiple data sources, and spatial misalignment arises when data are collected at different spatial scales and need to be incorporated into a single analysis. For instance, we may seek to understand the spatial distribution of cases at the state level when data were collected at the parish or county level (switching between two areal scales), or translate case data associated with household coordinates to cases at the county level (switching between point and areal scales) (Figure \ref{fig:fig1}C). Spatial big data has expanded the types of spatial information available for data aggregation \textemdash posts geo-tagged on social media might provide information at the level of countries, cities, neighborhoods, landmarks, and latitude-longitude coordinates \textemdash potentially engaging statistical change of support problems, even for one individual in a single day \cite{Gotway2002}. The multiplicity of highly resolved spatial scales also poses concerns for standard data checks, since traditional public health data will not necessarily be available at scales appropriate for validating comparisons to spatial big data \cite{Viboud2014, Shaman2013}. Finally, choices about how to deal with spatial misalignment have consequences for modeling results. For instance, recent studies have asked whether Zika virus-associated microcephaly was occurring at unusually high rates in different Brazilian states. Birth rate data might be collected at one spatial scale according to regular demographic surveys, but data systems tracking microcephalic live births would likely have finer spatial detail. Depending on the choice of spatial scale, the combination of these two data sources creates the potential for both over and under estimation of microcephaly rates.

\subsection*{Spatial confidentiality and ethics} 
The practice of collecting data without seeking appropriate ethical approval presents some risk for digital infectious disease epidemiology, and the access to fine-grain spatial information further deepens this concern. \edit{Safeguards currently implemented for collecting and sharing spatial big data have focused on the obfuscation and aggregation of shared data to protect privacy, and the anonymization and de-identification of individuals.} \edit{Many research institutions have standardized practices to protect individual privacy that follow the guidance of institutional review boards (IRBs), disclosure review boards for public use data, and federal laws (e.g., the Health Insurance Portability and Accountability Act of 1996 (HIPAA) in the United States), but these organizations do not often recognize high resolution spatial data as a source that should be covered under human subjects protection policies \cite{NRCPanel2007}.} Several studies have provided examples where seemingly anonymized data could be mined (or linked with other databases) for de-anonymization: de Montjoye et al. \cite{DeMontjoye2013a} showed that four spatio-temporal position points from mobile phone records can be sufficient to uniquely identify 95\% of individuals in a large de-identified dataset; and Homer et al. \cite{Homer2008} showed that the sheer quantity of data collected could be sufficient to re-identify individuals in a genetic database. These issues already push the limits of existing ethical review mechanisms and our understanding of de-anonymization. In the future, guidelines to protect privacy and confidentiality may require: the masking of individual-level records through the aggregation of data to coarser spatial resolutions (Figure \ref{fig:fig1}D), the provision of synthetic datasets that attempt to mimic underlying distributions \cite{Kinney2011}, or the distillation of spatial big data to parameters commonly used in epidemiological models. Investigations may consider the optimal choice of spatial scale in the context of trade-offs between the accurate representation of process heterogeneity and the protection of privacy \cite{NRCPanel2007} and the improvement of computational efficiency \cite{Deeth2016}. \edit{Nevertheless, public data become increasingly vulnerable to breaches of privacy as additional data are released and data mining techniques improve over time}.

\section*{Implications for public health communication and policy}

The promise of high spatial and temporal resolutions in spatial big data opens opportunities for change in the standard practice of public health. In circumstances where adjacent or subordinate administrative units issue separate public health recommendations (e.g., United States federal, state, and local governments may issue independent flu vaccination recommendations), spatial big data may enable these entities to derive their policies from analyses of a common dataset and encourage coordination of preparedness activities \edit{across scales} \citep[e.g.,][]{Lee2015}. \edit{There is a growing panoply of adaptive, behavioral, and health economic modeling methods aimed at identifying the most effective interventions for human and livestock diseases.} As these methods begin to find use during ongoing outbreaks, the combination of spatial big data and adaptive models could enable the real-time adaptive management of infectious diseases and the coordination of disease control efforts across spatial scales.  

In the long term, \edit{some sources of} big data \edit{may} become more readily available at finer spatial resolutions than the administrative regions at which policy decisions are made, \edit{even to the level} of the individual. Spatial big data has already changed consumer marketing strategies; rather than targeting geographic areas with certain socioeconomic and behavioral characteristics, marketers can now target individual users based on the behaviors demonstrated in their digital traces \cite{Dalton2015}. Should epidemiological modeling and design reflect these cultural changes to public health data? Perhaps an analogous scenario would see individual epidemiological data being used to inform optimal intervention strategies, \edit{ignoring the administrative boundaries that typically constrain decision making.} It is difficult to imagine how such a public health infrastructure could operate \textemdash resources must still be coordinated and expended by administrative units, and policy decisions must still apply to populations (rather than individuals) to maintain feasibility. Nevertheless, epidemiological analyses with spatial big data expand the possibilities for multi-scale coordination of infectious disease surveillance, response, and forecasting. 

The real-time high volume nature of spatial big data makes more epidemiological information readily available to policymakers, but it also creates challenges for the communication of public health information. Spatial big data enables small area analyses, which are simultaneously highly precise to spatial locations and highly uncertain in modeling results about risk of disease. Similarly, the rise of epidemic forecasting technologies based on spatial big data might present predictions about risk and epidemic outcomes in precise locations even though the forecasts themselves are subject to uncertainty \cite{Shaman2013}. Consumers of analyses derived from spatial big data \textemdash clinicians, public health officials, epidemiologists, modelers \textemdash should develop conscientious practices for communicating uncertainty about spatial results to the public.

\clearpage

\clearpage
\section*{Footnote page}

\subsection*{Funding}
This work was supported by the Jayne Koskinas Ted Giovanis Foundation for Health and Policy (JKTG) [dissertation support grant to ECL]; the National Cancer Institute at the National Institutes of Health [grant number R01CA172805 to ABL]; the Fogarty International Center at the National Institutes of Health; and the Research and Policy for Infectious Disease Dynamics program of the Science and Technology Directorate at the Department of Homeland Security. The opinions, findings, and conclusions or recommendations expressed in this material are those of the author and not necessarily those of JKTG, its directors, officers, or staff.

\subsection*{Acknowledgments}
The authors thank Shashank Khandelwal and two anonymous reviewers for their careful comments on earlier drafts of this work.

\subsection*{Conflicts of interest}
The authors declare no conflicts of interest.

\subsection*{Corresponding author}
Shweta Bansal\\
Assistant Professor, Department of Biology\\
Georgetown University\\
shweta.bansal@georgetown.edu; 202-687-9256\\
Elizabeth C. Lee (alternate)\\
Georgetown University\\
ecl48@georgetown.edu

\clearpage

\renewcommand\refname{References}
\bibliographystyle{vancouver}
\bibliography{spatial_nourl_v2.bib}

\end{document}